\documentclass[titlepage,12pt]{utarticle}
\usepackage{epsf}
\usepackage{amsmath,amsfonts}
\usepackage{array}
\usepackage[nosort]{cite}
\long\def\omit#1{}

\numberwithin{equation}{section}
\begin{document}

\preprint{
UTTG--26--97\\
{\tt hep-th/9712049}\\
}

\title{Aspects of ALE Matrix Models and Twisted Matrix Strings}

\author{David Berenstein, Richard Corrado, and Jacques Distler
        \thanks{Research supported in part by the Robert A.\ Welch
        Foundation and NSF Grant PHY~9511632.}}
\oneaddress{ Theory Group, Department of Physics\\
        University of Texas at Austin\\
        Austin TX 78712 USA \\ 
\email{david@zippy.ph.utexas.edu}
\email{rcorrado@zippy.ph.utexas.edu}
\email{distler@golem.ph.utexas.edu}
 }

\date{December 4, 1997}

\Abstract{We examine several aspects of the formulation of
M(atrix)-Theory on ALE spaces. We argue for the
existence of massless vector multiplets in the resolved $A_{n-1}$
spaces, as required by enhanced gauge symmetry in M-Theory, and 
that these states might have the correct gravitational
interactions. We propose a matrix model which describes M-Theory on 
an ALE space in the presence of wrapped membranes. We also consider 
orbifold descriptions of matrix string theories, as well as more
exotic orbifolds of these models, and present a classification
of twisted matrix string theories according to Reid's exact 
sequences of surface quotient singularities.} 

\maketitle
%%%%%%%%%%%%%%%%%%%%%%%%%%%%%%%%%%%%%%%%%%%%%%%%%%%%%%%%%%%%%%%%%%%%%

%%%
\renewcommand{\baselinestretch}{1.25} \normalsize

\section{Introduction}

Over the last year, a great deal of evidence supporting the
M(atrix)-Theory~\cite{BFSS:Conjecture} description of M-Theory has
been accumulated ({\em c.f.}, \cite{Banks:Matrix-Theory}
and references therein). An area which seems more or less well
understood is toroidal 
compactification, at least for tori of small dimension. The SYM on 
the dual
torus~\cite{BFSS:Conjecture,Taylor:Compact,Ganor:BranesFluxes,%
Berenstein:Moduli-Matrix}
description works well for $T^d$, $d\leq 3$, where the SYM is
renormalizable.  For $T^4$ and beyond, 
the resulting SYM is non-renormalizable, so a sensible definition 
must be given for the theory. Matrix descriptions of M-Theory
compactified on $T^4$ and $T^5$ have been
described~\cite{Rozali:Uduality,Berkooz-Rozali:FiveBrane,%
Seiberg:NewTheories}
in terms of the interacting six-dimensional theories with $(0,2)$
supersymmetry~\cite{Witten:variousdims,Strominger:OpenPBranes} and
attempts have been made to formulate similar models for
$T^6$~\cite{Losev:MnMs,Brunner:SixTorus,Hanany:SixandKK,%
Ganor:SixTorus}.
These issues have recently been reviewed
in~\cite{Sen:matrix-tori,Seiberg:WhyCorrect}. 

The study of M(atrix)-Theory on curved manifolds is also extremely
important. One particular case is ``compactification'' on an ALE
space, which captures many of the interesting features of K3
compactifications of M-Theory. In the spirit of the original
M(atrix) conjecture, this theory appears to be described by the 
theory of D0-brane partons moving on the ALE
space~\cite{Douglas:EnhancedMat,Douglas:Issues,%
Fischler:MatString-K3,Douglas:Strings97}.

The aim of this paper is to provide more evidence for the consistency
of the description of ALE compactifications of M-theory via ALE 
matrix models. These models are very different in spirit from the 
matrix model for compactification on $\text{K3}\times S^1$ that was 
proposed
in~\cite{Govindarajan:Note-Matrix,Berkooz:String-Dualities}. In
particular, in the case of the ALE models, the curved space is
represented by the moduli space of flat directions in the target
space, whereas in the $(0,2)$ model
of~\cite{Govindarajan:Note-Matrix,Berkooz:String-Dualities}, the K3
forms the base space of the theory. As yet, there is not much in the
way of a connection between these two descriptions. 

The outline of the paper is as follows. In section~\ref{sec:ale}, we
review the definition of ALE matrix models, considering the case of
the $A_{n-1}$ series in some detail. We then go on in
section~\ref{sec:vectors} to examine the
realization of enhanced gauge symmetry within the matrix models and
provide evidence in the $A_{n-1}$ matrix model for the existence of 
the spacetime vector multiplets  which remain massless in the 
blow-up, which are required for consistency with M-Theory. 

In section~\ref{sec:wrapped}, we discuss the existence and properties 
of wrapped membranes in the ALE matrix model, following 
Douglas~\cite{Douglas:EnhancedMat}. We claim that the physics of the 
wrapped membranes is described by a quiver gauge theory that 
corresponds to a particular pattern of gauge symmetry breaking in the 
standard ALE matrix model. As evidence for our proposal, we show that 
states exist which are supersymmetric vacua of the interacting 
(internal) part of the theory, but that the ground state energy of 
the decoupled $U(1)$ part of the theory (which corresponds to the 
motion of the center of mass in the five transverse flat dimensions) 
is non-zero. The dependence of the ground state energy on the D-term 
coefficients is that required for a BPS-saturated (massive) wrapped 
membrane state and the 16-fold degeneracy of the (non-supersymmetric) 
ground state yields the states of the seven-dimensional vector 
multiplet. We also find the expected Coulomb potential between 
membranes and antimembranes.

Further evidence for the existence of the massless vector multiplet 
states is provided in section~\ref{sec:orbifold}, where we consider
orbifold realizations of the ALE matrix quantum mechanics and matrix
string theories. We also discuss these orbifold realizations in the 
context of Witten's ``new'' gauge theories~\cite{Witten:NewGauge} 
and, in section~\ref{sec:classification}, we find that the 
types of matrix models that one can produce by orbifolding occur
according to Reid's classification of exact sequences 
of surface quotient singularities. In 
section~\ref{sec:dynamics}, we consider the dynamics of the massless
vectors and argue that it is plausible that they have the correct
gravitational interactions. 

\section{M(atrix)-Theory on ALE Spaces}
\label{sec:ale}

The construction of ALE matrix models is based on the hyperk\"ahler
quotient construction of supersymmetric sigma models with ALE target
spaces~\cite{Hitchin:Hyperkahler,Kronheimer:ALE}, as applied to
D-brane effective worldvolume
theories~\cite{Douglas:Quivers,Polchinski:Tensors-K3,%
Johnson:IIB-ALE}.

These models have their field content summarized by a
quiver diagram representing the extended Dynkin diagram of one of the
$A$-$D$-$E$ Lie algebras. To each vertex is associated the group 
$U(Nk_i)$, where $k_i$ is the Dynkin label of the $i$th vertex. In 
the field theory, the vertices are associated with six-dimensional 
vector multiplets, each of which transforms as the adjoint of the 
gauge group associated to the vertex, and as a singlet under the 
other groups. The edges of the quiver describe six-dimensional 
hypermultiplets that transform in the fundamental--anti-fundamental 
representations of the neighboring gauge groups, and as singlets of 
the other groups.

Matrix models are obtained from these gauge theories by a
dimensional reduction of these theories to $0+1$
dimensions. The large $N$ limit of the quantum mechanics should 
describe the infinite momentum frame limit of M-Theory on the ALE 
space, while the finite $N$ QM is conjectured to describe the 
discrete light-cone quantization (DLCQ) of 
M-theory~\cite{Susskind:AnotherConj}. In a similar fashion, the
dimensional reduction of the quantum mechanics to the
$1+1$-dimensional theory with base 
$S^1\times\BR$~\cite{Taylor:Compact}
describes IIA string theory at finite
coupling~\cite{Motl:Proposals,Banks:StringsMat,%
Dijkgraaf:Matrix-Strings},
as required by M-Theory--IIA
duality~\cite{Townsend:revisited,Witten:variousdims}.   

\subsection{The $A_{n-1}$ Series}

\begin{figure}[b]
\centerline{\epsfxsize=10cm \epsfbox{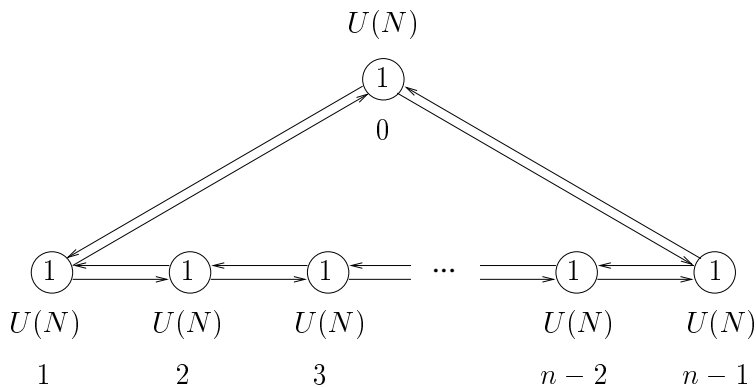}}
\caption{The $A_{n-1}$ quiver diagram.} \label{fig:a-series}
\end{figure}

Let us consider the explicit construction of the $A_{n-1}$ series. 
The
quiver diagram is shown in Figure~\ref{fig:a-series}.
Since the $k_i=1$, for each of the $n$ vertices there is
a $U(N)$ gauge group and a vector multiplet, $V_i$, transforming in
the $(1,\ldots,1,\text{ad}(U(Nk_i)),1,\ldots,1)$, whose bosonic
content will be written as $(A_{\mu i}, a_i)$. For each edge, we have
a hypermultiplet $H_{i,i+1}$ in the 
$(1,\ldots,1,Nk_i,\overline{Nk}_{i+1},1,\ldots,1)$ representation,
whose chiral components have the bosonic content 
$(x_{i,i+1},y_{i,i+1})$.   

>From this field content, we write down the most general action
with flat K\"ahler metric and common gauge couplings. The allowed
deformations of the theory are the addition to the lagrangian of F 
and
D-terms for the diagonal $U(1)$ gauge fields,
\begin{equation}
\begin{split}
D_i &= |x_{i-1,i}|^2+|y_{i,i+1}|^2-|x_{i,i+1}|^2-|y_{i-1,i}|^2+d_i\\
F_i &= x_{i-1,i} y_{i-1,i} - y_{i,i+1}x_{i,i+1} + f_i,
\end{split} \label{eq:dandfterms}
\end{equation}
where$\sum f_i = \sum d_i =0$. In higher-dimensional field
theories, in particular those obtained by further toroidal 
compactification,
the addition of theta terms are also allowed. The
hyperk\"ahler quotient consists of projecting onto field
configurations which are gauge invariant under the diagonal $U(1)$s,
such that the F and D-terms vanish.
The gauge invariant complex coordinates are given by 
\begin{equation}
\begin{split}
u &=\prod x_{i,i+1} \\
v &= \prod y_{i,i+1} \\
w& =w_i=x_{i,i+1} y_{i,i+1},
\end{split}
\end{equation} 
where, from the vanishing of the F-terms in~\eqref{eq:dandfterms}, 
all
of the $w_i$ are the 
same, modulo constant shifts on the moduli space of flat directions. 
It
follows that $u$, $v$, and $w$ satisfy 
\begin{equation}
uv = P(w),
\end{equation}
where $P(w)$ is a monic polynomial of degree $n$.  We see that
deformations by the D and F-terms in~\eqref{eq:dandfterms} correspond
to the blowing up of the $A_{n-1}$ singularity. As is conventional, 
we
denote the ALE space by $\CM_{\vec{\zeta}}$, where the $\vec{\zeta}$ 
are
the blow-up parameters. 

The low energy physics is described by quantum mechanics on the 
moduli
space of flat directions. The Higgs branch corresponds
to~\cite{Douglas:EnhancedMat}  
\begin{equation}
(\BR^5\times \CM_{\vec{\zeta}})^N/S_N
\end{equation}
where the $\BR^5$ corresponds to the flat directions given by the
global $U(1)$ 
under which none of the hypermultiplets transform. This describes the 
motion of
$N$ D0-branes on $\BR^7\times \CM_{\vec{\zeta}}$, and by the M(atrix)
conjecture describes M-theory on $\BR^7\times \CM_{\vec{\zeta}}$ in
the infinite momentum frame.  If we compactify one of the transverse
coordinates to $S^1$, we obtain a $1+1$ dimensional field theory with 
\begin{equation}
(\BR^4\times S^1\times\CM_{\vec{\zeta}})^N/S_N
\end{equation}
as its Higgs branch, which describes the IIA string theory on the ALE 
space.

\section{Massless Vector Multiplets in the Blow-up}
\label{sec:vectors}

Various issues concerning the ALE models are at hand. First, it is
necessary to provide the full  
massless spectrum from the quantum mechanics of the blown-up space, 
as
this must agree with the degrees of freedom expected from
supergravity considerations. On the other hand, new massless states
are expected to appear when the singularity is blown-down. These
states are visible from the viewpoint of the
ALE space as a description close to the singular point of the
degeneration limit of a large K3
surface~\cite{Witten:Strings95}. M[K3] is dual to
Het[$T^3$]~\cite{Witten:variousdims} and the 
degeneration limit we are considering is a point in the moduli space
with enhanced gauge symmetry (see~\cite{Aspinwall:TASI} and 
references
therein). The blow-up modes correspond to Higgsing
away these enhanced gauge groups, and so they form  part of a vector
multiplet in 7-dimensional physics. 

For the case of the compact K3, these states all arise from
2-branes wrapped around the 22~homology 2-cycles of the K3. When a
2-cycle shrinks to zero-size, the corresponding state becomes 
massless
and the gauge symmetry is enhanced. In the case of the ALE spaces,
the non-compactness of the space modifies the analysis slightly. For
example, in the case of the $A_{n-1}$ singularity, the
contribution from homology generates $n(n-1)$ states in the root
lattice of the enhanced gauge group. These states form 7-dimensional
vector multiplets and become massive when the singularity is
blown-up. They may be identified with bound states of 2-branes
wrapping the $\BP^1$s of the blow-up. There are an additional $n-1$
states in the Cartan subalgebra which are localized at the 
singularity
and remain massless in 
the blow-up. Additionally, there is a massless singlet state
arising from the self-dual cohomology of
the ALE. Since the SD form does not have compact support, the
wavefunction of this state is not normalizable. The Cartan
and singlet modes form massless 7-dimensional vector multiplets, so
the enhanced symmetry group is $U(n)$. 

All of these states should appear in the quantum
mechanics~\cite{Douglas:EnhancedMat,Douglas:Issues,Diaconescu:Matrix-
Mirror}. 
In the following, we will use a localization argument to show that
the states in the Cartan subalgebra exist and are localized near the
singularity. We provide 
further evidence that these states are normalizable ground states of
the quantum mechanics, and that there are exactly $n$ of them for a
single D0-brane moving on an $A_{n-1}$ ALE space.  

The number of massless vacua can be calculated in the following
manner. We begin by removing the
decoupled $U(1)$ and consider the case of a single D0-brane. Now, we
can deform the theory by adding hypermultiplet mass terms. Without
loss of generality, we can give the same mass to all  
of the hypermultiplets\footnote{Any other combination of
mass terms can be put into this form by shifting the values of the 
$U(1)$
chiral multiplets by a constant.}. This preserves $\CN=1$
supersymmetry and lifts 
the moduli space of vacua to a discrete set of points that solve  the
constraints in~\eqref{eq:dandfterms}. The remaining F-terms for the
diagonal $U(1)$ which are
required to vanish are 
\begin{equation}
\begin{split}
x_{i,i+1}(a_i- a_{i+1} + m) &= 0 \\
y_{i,i+1}(a_i-a_{i+1}+ m) & = 0.
\end{split} \label{eq:chargedchiral}
\end{equation}

The constraints~\eqref{eq:chargedchiral} require that the different
$U(1)$ chiral multiplets in the blown-up ALE space acquire
expectation values related to the mass perturbation. Only $n-1$ of
these F-terms can be set to zero in this manner for generic values of
the $d_i$ and $f_i$. This constrains the last chiral field to be set
to zero, which is the special point $u=v=0$ in the moduli space of
flat directions. Since there are exactly $n$ roots of  
the polynomial $P(w)=0$, there are $n$ vacua. In the presence of the
mass deformations, these vacua are normalizable, as there are no
non-compact flat directions. 

>From the M-Theory considerations discussed previously, one expects
that $n-1$ of these vacua should survive as normalizable states when
the mass perturbation is set to zero. While it would certainly be
advantageous to directly compute the Witten  
index in the non-compact ALE space via similar methods as those
in~\cite{Sethi:Redux,Porrati:BoundThreshold}, this is bound to be a 
complicated process, as it must be done in terms of the projected
coordinates, for which the action is quite non-linear.  

In order to consider the normalizability of these bound
states as the mass deformations are turned off, we will turn to the
dual picture of the ALE matrix model as the strong-coupling limit of
$n$ D6-branes in the IIA theory~\cite{Hanany:SixandKK}. In that
system, there are indeed $n$ states 
arising from the Cartan subalgebra of $U(n)$, but one of these states
corresponds to center-of-mass motion. We do not expect this state to 
be normalizable, but the remaining $n-1$ states which describe the 
relative motion of the D6-branes are. Therefore, as $m\rightarrow 0$ 
in the ALE matrix model, $n-1$ normalizable states survive and there 
is one non-normalizable state, which corresponds to the non-
normalizable singlet in the M-Theory picture that we discussed above.

The above counting of $n-1$ normalizable states agrees with the
familiar connection between supersymmetric quantum mechanics and
differential topology~\cite{Witten:SUSY-Morse}. As the supersymmetry
algebra has a representation in terms of the deRham cohomology on the
space of fields, we obtain $n-1$ massless states from the $n-1$
anti-self-dual forms on the $A_{n-1}$ ALE space. These forms have
compact support, so that the corresponding states are normalizable. 
We also obtain a single non-normalizable state from the self-dual
form. Similar considerations for the $D$ and $E$ series quantum 
mechanics would indicate that they should also possess the same 
states as in the M-Theory construction. However, our mass deformation
argument fails for the $D$ and $E$ series. In those cases, one can 
always change the values of the trace part of each of the $a_i$ to 
compensate for the mass term. The F-terms~\eqref{eq:chargedchiral} 
will not force us onto any special points in the moduli space of flat 
directions and we do not obtain any information on the counting of 
ground states. Without some version of an index theorem, we must be 
cautious about drawing conclusions with regard to the $D$ and $E$ 
series.

We note that the same deformations can be made in the $1+1$ matrix
SYM. As the spatial circle is compact, the higher oscillator modes
don't contribute to the index, so that a calculation of the index 
will again return the same number of ground states.

Now, for the normalizable vacuum states we found for the $A_{n-1}$
series, the decoupled quantum mechanics (corresponding to the
center-of-mass motion that we removed above) has 8 fermionic 
zero modes, where  four act as creation operators and four act as
destruction operators, giving a $2^4$ degeneracy of states. These are
exactly the states that form a massless vector multiplet in
seven dimensions. Also, the
different spaces that one obtains for higher-dimensional field
theories will have the appropriate number of moduli. One obtains 3 
from the D and F-terms (as corresponds to a 7-dimensional vector
multiplet), and when the matrix theory is further
compactified on a $d$-torus, there are
an additional $d$ theta angles~\cite{Berenstein:Moduli-Matrix},
\begin{equation}
\theta_i^a \int F_{0i}^a,
\end{equation}
where $\sum_{a} \theta^a_i=0$. 

On the other hand, the spectrum of states is continuous for the 
non-compact directions which are transverse to the ALE. These states
are non-normalizable and are 
acted on by supersymmetry, giving 8 extra fermionic zero modes. In
the spacetime picture, these states have a $2^8$ degeneracy, so that
they are identified as gravitons that propagate in the ALE
spacetime. This is expected from supergravity 
considerations, since asymptotically the space is flat. The
supersymmetry breaking occurs locally around the singularity, but far 
away from the origin, the SUSY is effectively restored.

As we see, the spectrum of low energy states for a single D0-brane
already contains all of the states  expected from
supergravity considerations. This also agrees with the following
argument in the $1+1$ dynamics. In the far IR limit, the gauge
coupling flows off to infinity and we recover a SCFT on the moduli
space of flat directions, namely matrix string
theory~\cite{Dijkgraaf:Matrix-Strings} on the ALE space. The twisted
sector long 
strings are interpreted as bound states of once-wound strings. In the
conformal field theory limit these sectors
are decoupled. Moreover, the central charge of the SCFT in each long
string is 12, as is expected for the light-cone degrees of freedom of
the type IIA string. Hence, the model already has all of the infrared
massless fields that are allowed. In particular, the blow-up modes of
the ALE space must already exist in the SCFT. By construction, the
projected variables that we are using in 
the field theory are gauge invariant, so the twisted
sectors of the SCFT orbifold should already exist as states in the
model and the blowing up of the singularity is manifest in these
variables. This also suggests that the above argument of $n-1$ 
normalizable vacua is also correct, as there are $n-1$ twisted 
sectors in the 
free string limit. Moreover, it is reasonable to assume that one has
bound states of $N$ of these strings that are stable even when one
relaxes the infrared limit,  as these are identified with the $S_N$
twisted sectors of the model. These long strings can attach 
themselves to the center of the ALE space, as argued previously, so 
it is reasonable to assume that these bound states 
exist in the full theory, {\em i.e.,} these states can carry
arbitrary longitudinal  momentum.

\section{Wrapped Membrane States}
\label{sec:wrapped}

Douglas~\cite{Douglas:EnhancedMat} has proposed that the states 
corresponding to wrapped membranes in the matrix model are the 
``fractional branes'' in the IIA description of D-branes at the 
orbifold point of an ALE space. Some of the features of these states 
were discussed in~\cite{Douglas:Issues,Douglas:Hard}. We would like 
to further investigate these wrapped membrane states in the matrix 
model. As we shall see, the gauge theory describing a wrapped 
membrane is a modification of the usual ALE matrix model. This gives 
a systematic prescription for performing calculations with wrapped 
membranes in the ALE matrix models.

Given the Kronheimer construction of the ALE space, one expects that 
a membrane wrapped around a $\BP^1$ is associated with the 
corresponding root of the extended Dynkin diagram. One can also have 
bound states of membranes wrapped around different
$\BP^1$, subject to the constraint that the the sum of all of the 
$\BP^1$s, weighted by the Dynkin indices, $k_i$, of the corresponding 
nodes of the extended Dynkin diagram, is homologically trivial. (The 
$A_1$ ALE space is a slightly degenerate case of this, as the two 
nodes of the Dynkin diagram correspond to the \emph{same} $\BP^1$ 
with opposite orientation. There are no bound states of wrapped 
membranes in this case.) 

Instead of the standard ALE matrix model with a $U(N k_i)$ gauge 
group associated to each node of the extended Dynkin diagram, we 
consider a more general quiver gauge theory with a 
$U(N_i)=U(N k_i+r_i)$ gauge group at each vertex, and hypermultiplets 
in the $(N_i,\bar{N}_j)$ associated to each link $<i,j>$.   As we 
shall see, this gauge theory describes $N$ D0-branes propagating on a 
ALE space with $r_i$ membranes wrapped around the $i$th $\BP^1$. 
After examining some of the features of this quiver gauge theory, we 
will see how it can be embedded in the standard ALE matrix model. 

As a first check, we see that, if all of the $r_i=nk_i$ for some 
integer $n$, the membranes are wrapped around a homologically trivial 
cycle and hence can be unwrapped. The configuration decays to D0-
branes and, indeed, from the point of view of the gauge theory, is 
equivalent to shifting the number of D0 branes, $N\to N+n$. 

Still, in the gauge theory with $N_i=N k_i$, even though there are no 
BPS-saturated configurations corresponding to wrapped membranes, 
there may be excited states of the gauge theory corresponding to 
asymptotically-separated membrane-antimembrane pairs. Indeed, this 
was the approach of Douglas 
{\it et.~al.}~\cite{Douglas:EnhancedMat,Douglas:Issues,Douglas:Hard}. 

More generally, when $\sum_i r_i\alpha_i$ is a root, we expect to 
find a (16-fold degenerate) ground state of the quantum mechanics, 
corresponding to the bound state of the corresponding collection of 
wrapped membranes. For other values of $(r_0,\dots,r_r)$ (modulo 
$(k_0,\dots,k_r)$), we expect to find flat directions corresponding 
to the fact that the wrapped membranes can be separated.

The simplest case, which we consider in detail below, is that of the
wrapped membrane in the $A_1$ theory (which is not expected to form 
bound states). Consider the matrix model for a single wrapped 
membrane (one of the $r_i=1$, the rest equal to zero). The 
Hamiltonian for this model still has a decoupled $U(1)$ which, as 
usual, has flat directions corresponding to the motion of the center 
of mass in the five transverse flat dimensions. As we shall see, the 
ground state energy in this decoupled $U(1)$ theory is nonzero (for 
nonzero D and F-terms),  fermionic zero modes still give rise to the 
same 16-fold degeneracy of the ground state, as in the standard ALE 
matrix model. Though the ground state is not supersymmetric, the 
$U(1)$ theory is free -- no fields are charged under the diagonal 
$U(1)$ -- and hence we can compute the ground state energy reliably. 
As we shall see, its dependence on the D-term coefficients is just 
what is needed for a BPS-saturated (massive) wrapped membrane state.

A massive vector multiplet in 7 dimensions, like the massless one, 
has 16~propagating degrees of freedom. This degeneracy is already 
accounted for by the degeneracy of the (non-supersymmetric) ground 
state of the decoupled $U(1)$ theory. So the state in the internal 
part of the theory must a supersymmetric ground state, annihilated by 
all of the supercharges.

To be slightly more general, let us consider the $A_1$ model  with 
gauge group $U(N)\times U(N+w)$, corresponding to $w$ wrapped 
membranes. We assume that $F=0$, so that only the D-terms,  
$D_{1,2}$, are non-zero. We can obtain a bound on the energy of the 
wrapped membrane state by rewriting the Hamiltonian in terms of the 
linear combinations of the above D-terms corresponding to the 
diagonal and internal $U(1)$s. After computing traces, the relevant 
part of the Hamiltonian is 
\begin{equation}
\begin{split}
H_D &=  N D_1^2 + (N+w) D_2^2 
+ N \, \zeta D_1 - (N+w) \, \zeta D_2 \\
&= \tilde{D}_1^2 + \tilde{D}_2^2 
+ \sqrt{N} \, \zeta \tilde{D}_1 
- \sqrt{N+w}\, \zeta \tilde{D}_2,
\end{split}
\end{equation} 
where we have defined normalized variables 
$\tilde{D}_1= \sqrt{N}\, D_1$, $\tilde{D}_2 = \sqrt{N+w}\, D_2$.
Since the decoupled $U(1)$ is the sum of the $U(1)$s at each vertex, 
the corresponding D-term, $D_{\text{dec.}}$, is associated to the sum 
$D_1+D_2$, we have the orthonormal pair
\begin{equation}  
\begin{split}
D_{\text{dec.}} &= \frac{1}{\sqrt{2N+w}} \, \Bigl( \sqrt{N}\, 
\tilde{D}_1 
+ \sqrt{N+w} \, \tilde{D_2} \Bigr) \\
D_{\text{int.}} &= \frac{1}{\sqrt{2N+w}} \, \Bigl( \sqrt{N+w}\, 
\tilde{D}_1 
- \sqrt{N} \, \tilde{D_2} \Bigr),
\end{split}
\end{equation}
where $D_{\text{int.}}$ is the D-term for the ``internal'' 
(difference) $U(1)$. Now the component of the Hamiltonian which 
depends on $D_{\text{dec.}}$ is
\begin{equation}
H_{\text{dec.}} = D_{\text{dec.}} 
\left( D_{\text{dec.}} - \frac{w\zeta}{\sqrt{2N+w}} \right),
\end{equation} 
which gives a bound 
\begin{equation}
E \geq \frac{w^2 \zeta^2}{4(2 N + w)} \label{eq:diag-bound}
\end{equation}
on the Hamiltonian.  Up to numerical factors, this is precisely what 
one expects for the energy of a membrane which is wrapped $w$ times 
around a sphere and has longitudinal momentum $N+w/2$. The mass of 
the membrane depends linearly on the blow-up parameter, $\zeta$, and 
is therefore proportional to the area of the sphere, as required. 

Now, in the case of the singly-wrapped membrane, $w=1$, let us show 
that the internal part of the theory is in a supersymmetric ground 
state and that there is a mass gap. Let the hypermultiplets be 
$(x_{12},y_{12})$ and $(x_{21}, y_{21})$, where $x_{12}$, $y_{21}$ 
are $N \times (N+1)$ and $x_{21}$, $y_{12}$ are 
$(N+1)\times N_1$ complex matrices.  We would like to minimize the 
Hamiltonian. We can keep $F=0$ by taking $x_{21}=y_{12}=0$, then we 
must minimize
\begin{equation}
\frac{1}{4} \, \text{tr}\left[ 
( x_{12} \bar{x}_{12} + y_{21} \bar{y}_{21} - \zeta )^2 
+ ( \bar{x}_{12}x_{12}  + \bar{y}_{21} y_{21}  - \zeta )^2 \right].
\label{eq:minimize}
\end{equation}
As the matrix operators $x_{12} \bar{x}_{12}$ and 
$\bar{x}_{12}x_{12}$ are positive, isospectral, and hermitian, we 
will take them to be diagonal and with ordered eigenvalues. The 
operator $\bar{x}_{12}x_{12}$ has one zero eigenvalue.  Similarly, 
$y_{21} \bar{y}_{21}$ and $\bar{y}_{21} y_{21}$ are isospectral.
Solution of the D-terms of the interacting piece will require that 
the sums 
\begin{equation}
\begin{split}
& x_{12} \bar{x}_{12} + y_{21} \bar{y}_{21} = A \cdot \Bid_{N} \\
& \bar{x}_{12}x_{12} + \bar{y}_{21} y_{21} = B \cdot \Bid_{N+1},
\end{split} \label{eq:diag-dterms}
\end{equation}
are proportional to the identity. Since the traces of these operators 
are equal, we must have $N A = (N+1)B$, so that~\eqref{eq:minimize} 
becomes
\begin{equation}
\frac{1}{4} \left[ \frac{N(2N+1)}{N+1} A^2 - 4N \zeta A + (2N+1) 
\zeta^2 \right].
\end{equation}
This is minimized by 
\begin{equation}
A = \frac{2\zeta(N+1)}{2N+1}
\end{equation}
and the bound obtained is
\begin{equation}
E= \frac{\zeta^2}{4 (2N+1)}, 
\end{equation}
which agrees with our earlier result~\eqref{eq:diag-bound}.

By again examining the trace of the sums in~\eqref{eq:diag-dterms} , 
we find that the eigenvalues of $x_{12} \bar{x}_{12}$ and 
$y_{21} \bar{y}_{21}$ are in arithmetic progression,
\begin{equation}
x_{12} \bar{x}_{12} = \text{diag}\left(\frac{N A}{N+1}, 
\frac{(N-1) A}{N+1},\ldots,\frac{A}{N+1}\right),
\end{equation}
while the eigenvalues of $y_{21} \bar{y}_{21}$ are in the opposite 
order. Similarly,
\begin{equation}
\bar{x}_{12}x_{12}  = \text{diag}\left(\frac{N A}{N+1}, 
\frac{(N-1) A}{N+1},\ldots,\frac{A}{N+1},0\right),
\end{equation}
with $\bar{y}_{21}y_{21}$ in the opposite ordering. This 
configuration breaks the interacting gauge group completely, so the 
states in the (gauge theory) vector multiplets are massive. 
Similarly, as the hypermultiplet appear squared in the non-vanishing 
D-terms, the hypermultiplets are also massive. Hence, the system has 
a mass gap, as expected.

Now, the above scheme for constructing a solution can fail for 
wrapping number $w=2$ and $N$ odd (so that we cannot construct a pair 
of branes separated from one another to satisfy the bound). In that 
case, the ranks of the matrices $x_{12} \bar{x}_{12}$ and 
$\bar{x}_{12}x_{12}$ will differ by 2 and they will both have an odd 
number of entries. Since $\bar{x}_{12}x_{12}$ now has two zero 
eigenvalues, we must find that all eigenvalues come in pairs. As the 
total number of eigenvalues is odd, the matrices on the left-hand 
side of~\eqref{eq:diag-dterms} can no longer both be proportional to 
the identity. This should be taken as evidence that two wrapped 
membranes do not form a bound state.  The results 
of~\cite{Sethi:Redux,Porrati:BoundThreshold} show that the $\CN=2$, 
$U(N)$ vector multiplet quantum mechanics does not have a bound 
state. One is therefore led to conjecture that the only bound states 
in our system will be those corresponding to the massive states in 
the adjoint of the enhanced gauge group, with an arbitrary number of 
D0-branes attached.

Now that we have seen some of the features of the wrapped membrane
matrix model, let us see how it can be recovered from a particular 
limit of the standard ALE matrix model. Consider a limit of the 
standard matrix model in which the $U(Nk_i)$ gauge symmetry at the 
$i^{th}$ node is broken to $(U(N_1 k_i+1)\times U(N_2 k_i -1) )$, 
with $N_1+N_2=N$. This corresponds to having a wrapped membrane and 
an anti-wrapped membrane around the $i^{th}$ $\BP^1$. Far out on the 
Coulomb branch, the membrane and anti-wrapped membrane are far apart 
in spacetime. The quantum corrections to the potential, obtained by 
integrating out the heavy strings which connect the membranes, vanish 
for large separation. So, in the limit in which the wrapped membrane 
and anti-wrapped membrane are infinitely far apart, the Hamiltonian 
splits into two pieces, each of which is of precisely the form of the 
matrix model we have proposed. More general gauge symmetry breaking 
patterns correspond to more complicated configurations of
wrapped and anti-wrapped membranes around various $\BP^1$s.

We will now use this formalism to compute the interaction between a  
membrane and an anti-wrapped membrane. For the $A_1$ case, consider 
the standard $U(N)\times U(N)$ quiver gauge theory, with the gauge 
symmetry broken to
\begin{equation}
( U(N_1)\times U(N_2) )\times ( U(N_1+1)\times U(N_2-1) ),
\label{eq:a1-broken}
\end{equation}
where $N_1 + N_2 = N$. Our notation is such that the $U(N)$ of each
vertex in the quiver diagram is broken to one of the factors in 
parentheses. While the membrane states that we have considered to 
this point have been associated with the $U(N_1)\times U(N_1+1)$ 
component of the gauge group~\eqref{eq:a1-broken}, the anti-wrapped 
membranes are associated with the $U(N_2)\times U(N_2-1)$ component. 
The crucial distinction is the difference in the VEVs taken in the 
$U(1)$s at the different vertices. The membrane--anti-wrapped
membrane solution occurs when all of the hypermultiplets vanish and
has mass squared proportional to 
$\bigl(M_P^{(7)}\bigr)^4 r^2 - \zeta$, where $r$ is the separation 
between the membranes. Integrating out the hypermultiplets in the 
quantum theory will induce a potential
\begin{equation}
\begin{split}
& 4 \sqrt{\bigl(M_P^{(7)}\bigr)^4 r^2 + \zeta} \, 
+ 4 \sqrt{\bigl(M_P^{(7)}\bigr)^4 r^2 - \zeta} \,
- 8 \bigl(M_P^{(7)}\bigr)^2 r \\
& \hspace*{2cm} \sim 
- \frac{2\zeta^2}{\bigl(M_P^{(7)}\bigr)^6 r^3} 
+  r O((\zeta/r^2)^3).
\end{split}
\end{equation}
This is the Coulomb potential expected for BPS objects whose mass and 
charge are proportional to $\zeta$. 

The results that we have presented above show that these states can 
carry an arbitrary amount of longitudinal momentum and seem to have 
the right properties for an interpretation as wrapped membrane 
states. The study of these states in more detail, as well as the 
massless states lying in the Cartan algebra that we discussed in 
section~\ref{sec:vectors}, seems very promising. In particular, the 
ground states describing the wrapped membranes seem to exhibit a very
interesting structure that should be exploited to extract more 
information about the structure of the ALE matrix models.

\section{Considerations from M(atrix) Orbifolds}
\label{sec:orbifold}

When one considers $\BC^2/\Gamma$ orbifold string theories, the
orbifolding procedure introduces new twisted sectors which serve to
restore the modular invariance of the partition function. 
Furthermore,
for $\Gamma=\BZ_n$, there arises a quantum $\BZ_n$ symmetry of the
twisted fields. Orbifolding with respect to this quantum symmetry
reproduces the original unorbifolded theory.

As the blow-up parameters transform non-trivially under the quantum
symmetry, blowing up the singularity explicitly breaks the quantum
symmetry. We can describe the blow-ups via the hyperk\"ahler
quotient construction of the $A_{n-1}$ ALE spaces. The 
quantum symmetry is always generated by the outer automorphisms
of the Lie algebra. For the $A_{n-1}$ case at hand, these permute the
roots in a fashion which is 
represented by clock-shifts on the extended Dynkin diagram in
Figure~\ref{fig:a-series}. In terms of the ALE matrix model, this
corresponds to a clock-shift on the vectors and hypermultiplets,
\begin{equation} 
\begin{split}
V_i & \rightarrow V_{i+k}\\ 
H_{i,i+1}  & \rightarrow H_{i+k,i+k+1}, 
\end{split} \label{eq:clockshifts}
\end{equation}
which leaves the action invariant. In this manner, the quantum 
symmetry also
acts on the F and D-terms by the same clock-shifts. 

Now the clock-shifts in~\eqref{eq:clockshifts} correspond to the
representations of $\BZ_n$ on the fields. The vacua we found also
transform into one another, via
\begin{equation}
|\phi_i> \to |\phi_{i+k}>,
\end{equation}
as each corresponds to which pair $(x_{i,i+1},y_{i,i+1})=0$. We note
that there is one  
state that transforms invariantly under the transformation, namely
$\sum_i |\phi_i>$. This is the singlet state discussed in
section~\ref{sec:vectors}. This correspondence provides further
evidence that the states that we have constructed are indeed the ones
corresponding to the Cartan subalgebra of $U(n)$. 

We note again that these considerations are independent of whether we
are working within the quantum mechanics or in the 1+1 field theory.
We therefore obtain consistent descriptions of both M-Theory and the 
IIA
string on the ALE space.

Recently, Witten~\cite{Witten:NewGauge} examined the new physics
which arises in certain exotic orbifolds of M-Theory, as well as a
matrix model description of a class of such models. Witten considers
M-theory on $(\BC^2\times S^1)/\Gamma$, for which it turns out that
one can obtain a gauge group in six dimensions whenever there 
exists an exact sequence 
\begin{equation}
0 \longrightarrow \Gamma^\prime \longrightarrow \Gamma
\longrightarrow \BZ_n \longrightarrow 0, \label{eq:exactsequence}
\end{equation}
for $\Gamma^\prime\subset\Gamma$ a finite subgroup of $SU(2)$ and 
some
cyclic group $\BZ_n$. A classification of such exact sequences can be
found in Reid~\cite{Reid:Young-Persons}. As $\Gamma$ acts 
transitively
on $S^1$ via the $\BZ_n$ action, one obtains a  
circle of $\Gamma^\prime$ singularities with $\BZ_n$ monodromies. The
monodromies act by outer automorphisms of $\Gamma^\prime$ which 
breaks
the $A$-$D$-$E$ group associated to 
$\Gamma^\prime$ to the visible gauge group. The spacetime singularity 
is 
$\BC^2/\Gamma^\prime$. 

It is interesting to consider similar types of constructions of 
matrix
theories. In particular, the various ways that one can twist the
boundary conditions will lead to different physics in the orbifold 
limit.

Let us consider the standard $1+1$-dimensional matrix theory. When
going around the $S^1$ of the matrix model base space, we can twist
the hypermultiplet by a $m$th root of unity, $\omega$, and leave the
vector multiplet alone,
\begin{equation}
\begin{split}
V(\sigma+2\pi) & = V(\sigma) \\ 
H(\sigma+2\pi) & = \omega H(\sigma). \label{eq:twistloop} 
\end{split} 
\end{equation}
The resulting boundary conditions on the hypermultiplet break half of
the supersymmetry and splits it into a pair of chiral multiplets. 

In the IR limit, we obtain some version of matrix string theory.
As the eigenvalues for the 
hypermultiplets must also satisfy the $\BZ_m$ identification
in~\eqref{eq:twistloop}, the moduli space for the particles is
$$\BR^4\times (\BC^2/\BZ_m).$$ 
So this is matrix string theory on an $A_{m-1}$ space. When
gluing fields together to make long strings, we find $m$ 
different twisted sectors classified by the length of the string 
modulo $m$. 
In particular, for all strings of length $k=  1,\dots, m-1\mod{m}$, 
the
hypermultiplets are not periodic and cannot have a zero-mode, so they 
are stuck to the zero locus for a supersymmetric vacuum.
On the other hand, for the strings whose length is a multiple of $m$, 
the hypermultiplets acquire a zero-mode, so that there is a Higgs
branch for these sectors. 

The above counting of states indicates that
the ground states of the different length strings yield $m-1$
six-dimensional vector multiplets. Ignoring the circle, these states
are localized at the origin of $\BC^2/\BZ_m$. By our mass deformation
arguments in section~\ref{sec:vectors}, we are also led to believe
that there is an extra bound state developed in the last sector. If
this is true, then when one extracts the weakly coupled IIA theory by
taking the radius of the circle to zero size, the spectrum of states
is precisely that required to have an $A_{m-1}$ ALE singularity. The
fractional strings become the twisted sectors of the SCFT. It is 
rather
important to notice that the splitting and joining of long strings
occurs according to the fusion rules of the orbifold SCFT. 

This construction should correspond to M-Theory on 
$(\BC^2\times S^1)/\BZ_m$, where the $\BZ_m$ acts transitively on
$S^1$. In terms of the exact
sequence~\eqref{eq:exactsequence}, $\Gamma^\prime$ is trivial and
there is no enhanced gauge symmetry in the six-dimensional physics. 

For the $A_{n-1}$ matrix models we can also twist one of the
hypermultiplets by an $m$th root of unity. In particular,
the gauge-invariant coordinates transform as 
\begin{equation}
\begin{split}
u & \rightarrow \omega u \\
v & \rightarrow \omega^{-1} v \\
w & \rightarrow w.
\end{split}
\end{equation}
The origin, $u=v=0$ is the only point left fixed by this
transformation, so the supersymmetric ground states are those for
which one of the hypermultiplets is set to zero. Since one can change 
which
of the hypermultiplets is shifted by large gauge transformations, 
there are $n$ such states, which are again vectors in
six-dimensions. Long strings are now classified by their congruence
modulo $m$, so that, in total, we find $nm$ vectors. This is
expected from an analysis of the CFT limit, as it corresponds to the
IIA string on the $\BC^2/\BZ_{nm}$ orbifold. As seen from the M-
Theory
perspective, the orbifold group is acting by $\BZ_m$ actions on the
shrunken circle. 

For the $D$ and $E$ series, this extra twisting can always be removed 
by a
large gauge transformation. Therefore, no new physics will arise by
such an orbifolding of the theory. 

Now, in Witten's matrix model~\cite{Witten:NewGauge}, the twisting is
done via the outer automorphisms of the Lie algebra, by clock-shifts
of size $p$. When one examines the moduli space of flat directions,
one sees that it parameterizes $\BC^2/\BZ_k$, where
$k=\text{gcd}(n,p)$. This model describes M-Theory
compactified on $(\BC^2\times S^1)/\BZ_n$, where $\BZ_n$ acts by
$\BZ_{n/k}$ actions on the $S^1$. Corresponding to the exact sequence
\begin{equation}
0 \longrightarrow \BZ_k \longrightarrow \BZ_n
\longrightarrow \BZ_{n/k} \longrightarrow 0, \label{eq:sequence-
witten}
\end{equation}
we have a circle of $\BC^2/\BZ_k$ singularities with a trivial 
$\BZ_{n/k}$
monodromy, so the spacetime gauge group is $U(k)$. 
On the other hand, we started with the $A_{n-1}$ matrix model, so we
should think of the end product as a circle of $\BZ_n$ singularities
with a $\BZ_{n/k}$ monodromy which generates $n/k$ images for each
shrunken $\BP^1$. 

Now, we see that there are several twisted models which yield $U(n)$
gauge groups. We have summarized the four models, as well as their, 
at
least tentative, M-Theory interpretation in 
Table~\ref{tab:a-series-models}.
\begin{table}[h]
\begin{center}
\begin{tabular}{|c|c|c|}
\hline
Matrix Model & Type of Twist & M-Theory Interpretation \\
\hline
standard & $\BZ_n$ phase on hypermultiplet 
& $(\BC^2\times S^1)/\BZ_n$ \\
\hline
$A_{n-1}$ & none & $\BC^2/\BZ_n \times S^1$ \\
\hline
$A_{n/m-1}$ & $\BZ_m$ phase on hypermultiplet 
& $(\BC^2/\BZ_{n/m} \times S^1)/\BZ_m$ \\
\hline
$A_{mn-1}$ & $\BZ_m$ clockshift & $(\BC^2/\BZ_{mn} \times S^1)/\BZ_m$
with monodromies \\
\hline
\end{tabular}
\end{center}
\caption{The four types of twisted matrix models with $U(n)$ gauge
group and their suggested M-Theory interpretations.}
\label{tab:a-series-models} 
\end{table}
According to Witten~\cite{Witten:NewGauge}, the feature which should
distinguish the vector theories at the singularity given by these
models is the theta angle in the six-dimensional gauge theory. For
example, the $A_{n-1}$ model has $\theta=0$, while the $A_{n/m-1}$
model twisted by a $\BZ_m$ phase on hypermultiplet and the
$A_{mn-1}$ model twisted by $\BZ_m$ clockshifts both have $\theta\neq 
0$.  

Finally, we note that we can allow more general types of twisted
models by combining the two types of twisting.

\section{Classification of Twisted Matrix Models}
\label{sec:classification}

In the previous section, we considered Witten's matrix model, which
amounted to orbifolding the $A_{n-1}$ matrix models by elements of 
their
quantum symmetry group. This is, in fact, one particular case of a
general construction of twisted matrix models based on Reid's exact
sequences~\eqref{eq:exactsequence}. In case~(1) of
Table~\ref{tab:reid}\footnote{As referenced in
Witten~\cite{Witten:NewGauge}, D.\ Morrison has pointed out that 
there
is a typo in case (2) on page~376
of~\cite{Reid:Young-Persons}. $D_{2n+1}$ should read $D_{2n+3}$, as 
appears correctly in Table~\ref{tab:reid}.}, the models based on
$\Gamma^\prime=\Gamma(A_{n-1})$ and $\Gamma=\Gamma(A_{rn-1})$ are the
second and fourth entries, respectively, in
Table~\ref{tab:a-series-models}.   
\begin{table}[h]
\begin{center}
\setlength{\extrarowheight}{2pt}
\begin{tabular}{|c|l|}
\hline
Case & $0 \longrightarrow \Gamma^\prime \longrightarrow \Gamma
\longrightarrow \BZ_n \longrightarrow 0$ \\
\hline
(1) & $0 \longrightarrow \Gamma(A_{n-1}) \longrightarrow
\Gamma(A_{rn-1}) \longrightarrow \BZ_r \longrightarrow 0$ \\
\hline
(2) & $0 \longrightarrow \Gamma(A_{2n}) \longrightarrow
\Gamma(D_{2n+3}) \longrightarrow \BZ_4 \longrightarrow 0$ \\  
\hline 
(3) & $0 \longrightarrow \Gamma(A_{2n-1}) \longrightarrow 
\Gamma(D_{n+2}) \longrightarrow \BZ_2 \longrightarrow 0$ \\
\hline
(4) & $0 \longrightarrow \Gamma(D_4) \longrightarrow \Gamma(E_6)
\longrightarrow  \BZ_3 \longrightarrow 0$ \\
\hline
(5) & $0 \longrightarrow \Gamma(D_{n+1}) \longrightarrow
\Gamma(D_{2n}) \longrightarrow  \BZ_2 \longrightarrow 0$ \\
\hline
(6) & $0 \longrightarrow \Gamma(E_6) \longrightarrow \Gamma(E_7)
\longrightarrow  \BZ_2 \longrightarrow 0$ \\
\hline
\end{tabular}
\setlength{\extrarowheight}{0pt}
\end{center}
\caption{Reid's six classes of surface quotient singularities.}
\label{tab:reid}
\end{table}
This is a degenerate
case, however, because the $\BZ_r$ action on 
$\Gamma(A_{n-1})$ is trivial, so the ``twisted'' theory
based on $\Gamma^\prime$ is just the $A_{n-1}$ ALE matrix model. On
the other hand, $\BZ_r$ acts by clockshifts on
$\Gamma(A_{rn-1})$, which yields Witten's twisted model. 

The generalization of this amounts to considering all possible
twistings by the symmetry groups of the $A$-$D$-$E$ extended Dynkin
diagram of an ALE matrix model. Since we are discussing twists when
going around a circle, we must restrict ourselves to orbifolding by
cyclic symmetries of the extended Dynkin diagrams. In this section, 
we
show that the most general twists that are allowed and which lead to
gauge groups are the twisting of $\Gamma^\prime$ and $\Gamma$ by the
cyclic groups $\BZ_n$ appearing in Reid's classification. 
The orbifolds generated lead to new matrix models and, hopefully, new
physics.   

\subsection{Models with $Sp(n)$ Gauge Group}

\begin{figure}[h]
\centerline{\epsfxsize=10cm \epsfbox{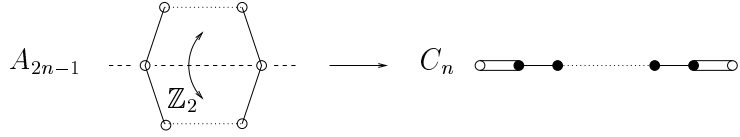}}
\caption{The $\BZ_2$ twist on the $A_{2n-1}$ model that generates a
$C_n$ model.} 
\label{fig:a2n-1-cn} 
\end{figure}

Let us consider Reid's case~(3). Here $\Gamma^\prime=\Gamma(A_{2n-
1})$ 
contains an even number of vertices, so that the $\BZ_2$ reflection 
on
$A_{2n-1}$ in Figure~\ref{fig:a2n-1-cn} yields a diagram which
resembles the extended Dynkin diagram for $C_n$.

The fields at the fixed vertex, call it the $i$th,  must transform as 
\begin{equation}
\begin{split}
V_i &\rightarrow V_i \\
x_{i-1,i} &\rightarrow y_{i,i+1} \\
y_{i-1,i} &\rightarrow -x_{i,i+1},
\end{split}
\end{equation}
so that the F-term is preserved, $F_i\rightarrow F_i$. However, this
seems to break half of the supersymmetry, since the chiral fields
$(x,y)$ no longer form a hypermultiplet. In particular, the
gauge-invariant coordinates transform as
\begin{equation}
\begin{split}
u &\rightarrow v \\
v &\rightarrow u \\
w &\rightarrow -w,
\end{split}
\end{equation}
which is a $\BZ_2$ action.

However, since there are an even number of vertices for the 
$A_{2n-1}$ diagram, we can place the hypermultiplets
in the $(1,\ldots,1,N k_i, N k_{i+1}, 1,\ldots,1)$ representations. 
In this case, one can easily see that these assignments actually 
preserve all of the supersymmetry. The new assignments result in 
different F and D-terms than appeared in~\eqref{eq:dandfterms}. 
Here we have
\begin{equation}
\begin{split}
D_i &= |x_{i-1,i}|^2-|y_{i,i+1}|^2+|x_{i,i+1}|^2-|y_{i-1,i}|^2 \\
F_i &= x_{i-1,i}y_{i-1,i} + y_{i,i+1} x_{i,i+1}.
\end{split}
\end{equation}
In particular, when orbifolding by symmetries of the extended Dynkin
diagram, the fields will transform as 
\begin{equation}
\begin{split}
V_i &\rightarrow V_{i^\prime} \\
(x_{i-1,i},y_{i-1,i}) &\rightarrow 
(x_{i^\prime-1,i^\prime},y_{i^\prime-1,i^\prime})  \\
D_i,F_i &\rightarrow D_{i^\prime},F_{i^\prime}, 
\end{split}
\end{equation}
thereby preserving all of the supersymmetry. 

\begin{figure}[h]
\centerline{\epsfxsize=14cm \epsfbox{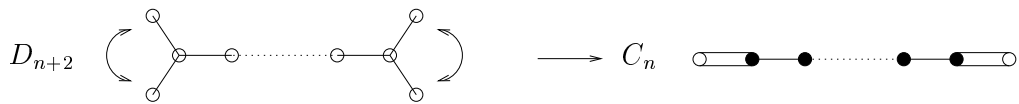}}
\caption{The $\BZ_2$ action on the $D_{2n}$ diagram induced by the
reflection of Figure~\ref{fig:a2n-1-cn}.}
\label{fig:d2n-z2} 
\end{figure}

Now let us consider $\Gamma=\Gamma(D_{n+2})$. As the $D$ and $E$ 
series are
described by open quivers, we can place the hypermultiplets in the
fundamental--fundamental representations, as we did for $A_{2n-1}$
above. We can therefore also preserve all of the supersymmetry in
the twisted $D$ and $E$ models.  From the exact sequence 
\begin{equation}
0 \longrightarrow \Gamma(A_{2n-1}) \longrightarrow \Gamma(D_{n+2})
\longrightarrow \BZ_2 \longrightarrow 0, 
\end{equation} 
we can tell which $\BZ_2$ action on the $D_{n+2}$ that the reflection
in Figure~\ref{fig:a2n-1-cn} induces\footnote{The induced actions for
Reid's cases (3)-(6) follow from the discussion of induced
representations found in Appendix~III of
\cite{Slodowy:Simple-Singularities}.}; it is the identification found
in Figure~\ref{fig:d2n-z2}.

To complete the identification of the new diagram obtained in
Figure~\ref{fig:a2n-1-cn} with the $C_n$ extended Dynkin diagram, we
need to define a consistent set of rules for obtaining the new roots
from the old roots. To determine these rules, we consider the
M-Theory interpretation of these models. Each root of the extended
Dynkin diagram is associated with a $\BP^1$ that can be blown-up in
the ALE space, and these $\BP^1$s are further associated to the 
wrapped
membrane states.  The roots, $\alpha_i$, are not linearly 
independent,
but satisfy the relationship 
\begin{equation}
\sum_i k_i \alpha_i =0, \label{eq:roots}
\end{equation}
where the $k_i$ denote the Dynkin labels\footnote{Or marks, if the
group is not simply-laced.}  of the extended algebra. In
terms of the $\BP^1$s, \eqref{eq:roots} implies that
the topological sum of the $\BP^1$s is trivial in integral homology.

Now, when we consider taking the $\BZ_n$ quotient of our $A$-$D$-$E$
matrix model, it is clear that we must require that the weighted sum
of the $\BP^1$s of the quotient is
trivial in the integral $\BZ_n$-equivariant homology.  
Therefore, our rules for determining the new roots, 
$\tilde{\alpha}_i$
and the metric on them for the twisted models must obtain the
condition~\eqref{eq:roots}
\begin{equation}
\sum_i \tilde{k}_i \tilde{\alpha}_i =0, \label{eq:newroots}
\end{equation}
where $\tilde{k}_i$ are the marks of the quotient diagram.

\begin{figure}[h]
\centerline{\epsfxsize=14cm \epsfbox{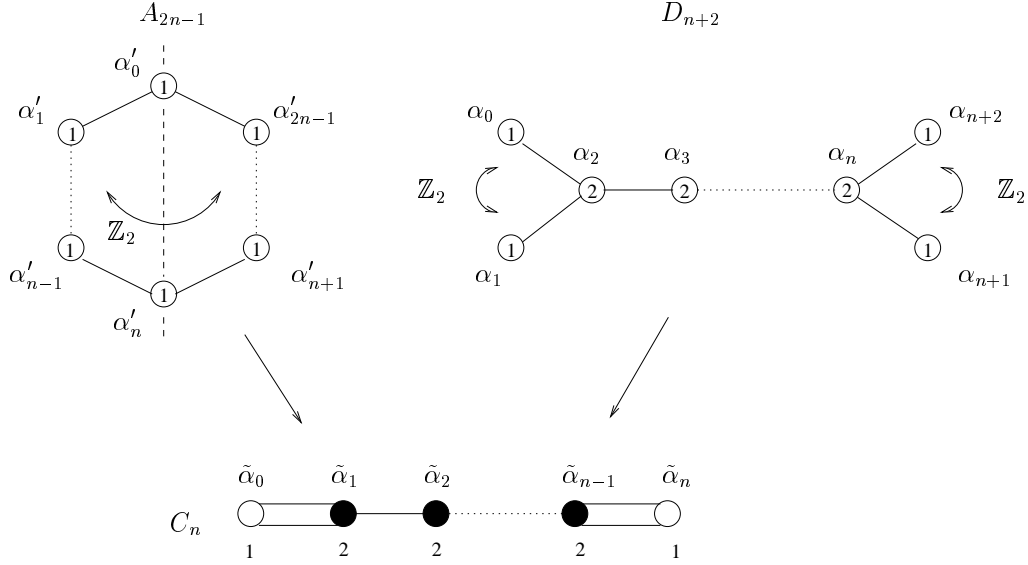}}
\caption{The labeling of roots and Dynkin labels (marks) for $A_{2n-
1}$,
$D_{n+2}$, and $C_n$.} \label{fig:a-d-c}
\end{figure}

Figure~\ref{fig:a-d-c} gives a labeling of the roots for the
$A_{2n-1}$, $D_{n+2}$, and $C_n$ models.
For $C_n$, the vanishing condition~\eqref{eq:newroots} is the sum 
over the
roots weighted by the marks,
\begin{equation}
\tilde{\alpha}_0 + 2 \sum_{i=1}^{n-1} \tilde{\alpha}_i+
\tilde{\alpha}_n = 0. \label{eq:roots-cn}
\end{equation}
This requires that we obtain new roots from the $\BZ_2$ quotient of
$\Gamma(A_{2n-1})$ according to
\begin{equation}
\begin{split}
\tilde{\alpha}_0 &= \alpha_0^\prime \\
\tilde{\alpha}_i &= \frac{1}{2} \left(\alpha_i^\prime +
\alpha_{2n-i}^\prime \right), ~~~i=1,\ldots,n-1 \\
\tilde{\alpha}_n &= \alpha_n^\prime. 
\end{split}
\end{equation} 
Now consider the new roots obtained after the $\BZ_2$ action on
$\Gamma(D_{n+2})$. in order for the new roots obtained from the 
$\BZ_2$
action on $\Gamma(D_{n+2})$ to satisfy~\eqref{eq:roots-cn}, they must
be given as
\begin{equation}
\begin{split}
\tilde{\alpha}_0 &= \alpha_0 + \alpha_1  \\
\tilde{\alpha}_i &= \alpha_{i+1}, ~~~~~i=1,\ldots,n-1 \\
\tilde{\alpha}_n &= \alpha_{n+1}+\alpha_{n+2}. 
\end{split}
\end{equation}
According to this prescription, these $\BZ_2$ orbifolds of
the $A_{2n-1}$ and $D_{n+2}$ ALE 
matrix models yield twisted matrix string theories whose
six-dimensional physics has an $Sp(n)$ gauge symmetry.

We find that, in general, when twisting the $\Gamma^\prime$ and
$\Gamma$ models, we must ensure that the sum of $\BP^1$s in the
quotient forms an integral class (and not a multiple of one) and that
at least one of 
the old roots with Dynkin label $k_i=1$ appear with coefficient one 
in
the expression for the new roots. We normalize the metric so that the
longest root has $(\text{length})^2=2$. The rules we must use to
obtain the new roots and metric on them are the following:
\begin{enumerate}
\item The new roots are determined from the old roots according to 
the
formula
\begin{equation}
\tilde{\alpha}_i = \frac{\text{min}(n_i,n_e)}{n_i}
\sum_{A_i} \alpha_{A_i}. \label{eq:ruleone}
\end{equation}
In this formula, $n_{i}$ denotes the number of old roots which are
pre-images of the new root $\tilde{\alpha}_i$, $A_i$ is the index set
which labels these pre-images, and $n_e=\text{min}\{n_i|k_i=1\}$.
\item The metric on the new roots is proportional to the induced
metric
\begin{equation}
\tilde{g}_{ij} = \frac{1}{n_e} \tilde{\alpha}_i \cdot 
\tilde{\alpha}_j.
\label{eq:ruletwo}
\end{equation}
\end{enumerate}

We note that in the case of the $\Gamma^\prime$ diagrams, the 
symmetry
we quotient by is present in the case of the {\em unextended} Dynkin
diagram, so that the extended root may be left fixed. In that case
$n_e=1$ and the rules 1 and 2 reduce to the prescription described by
Aspinwall and Gross~\cite{Aspinwall:so32-k3} in their consideration 
of
symmetries of the unextended Dynkin diagrams, namely
\begin{equation}
\Gamma^\prime: 
\begin{cases}
\tilde{\alpha}_i = \frac{1}{\text{\# of pre-images}} \sum
(\text{pre-images}) \\
\tilde{g}_{ij} = \tilde{\alpha}_i \cdot \tilde{\alpha}_j.
\end{cases}
\end{equation} 
In the case of the $\Gamma$ diagrams, the symmetry always acts on the
extended root, so that $n_e=\text{ord}(g)$, where $g$ is the 
generator
of the symmetry group, and
\begin{equation}
\Gamma: 
\begin{cases}
\tilde{\alpha}_i = \sum (\text{pre-images}) \\
\tilde{g}_{ij} = \frac{1}{\text{ord}(g)} \tilde{\alpha}_i \cdot
\tilde{\alpha}_j. 
\end{cases}
\end{equation} 

With a set of twisting rules in hand, let us revisit Reid's case~(1),
which contains Witten's matrix model. For
$\Gamma^\prime=\Gamma(A_{n-1})$, the $\BZ_r$ action is trivial, so 
the
new roots are exactly the same as the old roots. For
$\Gamma=\Gamma(A_{rn-1})$, $\BZ_r$ acts by a clockshift which yields
an $A_{n-1}$ diagram under identification of vertices. Using the
rules~\eqref{eq:ruleone} and~\eqref{eq:ruletwo}, we find
\begin{equation}
\tilde{\alpha}_i=\sum_{j=1}^r \alpha_{i+(r-j)n},
\end{equation}
so that
\begin{equation}
\tilde{\alpha}_i^2 =  \frac{1}{r} \sum_{k=1}^r \sum_{l=1}^r
\alpha_{i+(r-k)n} \cdot \alpha_{i+(r-l)n} =2
\end{equation}
and $\sum_i \tilde{\alpha}_i = 0$.
 
\subsection{Models with $G_2$, $SO(2n-1)$, and $F_4$ Gauge Symmetry}

\begin{figure}[h]
\centerline{\epsfxsize=10cm \epsfbox{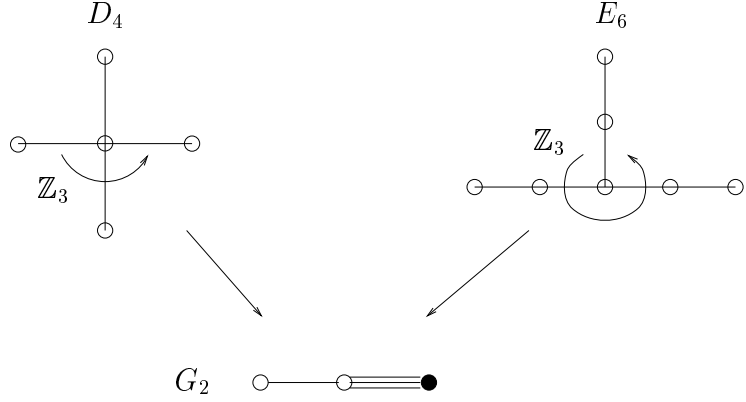}}
\caption{The $\BZ_3$-twisted $\Gamma(D_4)$ and $\Gamma(E_6)$ models
with $G_2$ gauge group.} \label{fig:d-e-g}
\end{figure}

For Reid's case~(4) in Figure~\ref{fig:d-e-g}, we again find a pair 
of
models which, after twisting according to the appropriate set of
rules, lead to a theory with a $G_2$ gauge group. 

\begin{figure}[h]
\centerline{\epsfxsize=10cm \epsfbox{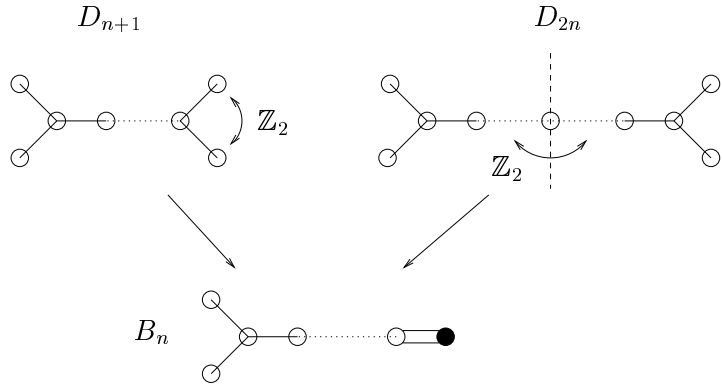}}
\caption{The $\BZ_2$-twisted $\Gamma(D_{n+1})$ and $\Gamma(D_{2n})$ 
models
with $SO(2n-1)$ gauge group.} \label{fig:d-d-b}
\end{figure}

\begin{figure}[h]
\centerline{\epsfxsize=10cm \epsfbox{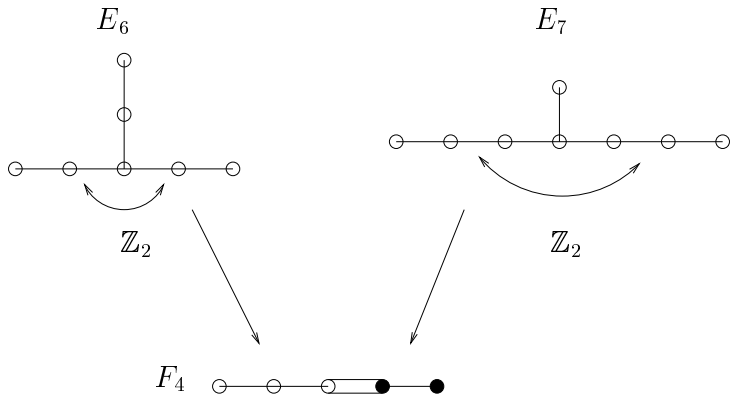}}
\caption{The $\BZ_2$-twisted $\Gamma(E_6)$ and $\Gamma(E_7)$ models
with $F_4$ gauge group.} \label{fig:e-e-f}
\end{figure}

Case~(5) in Figure~\ref{fig:d-d-b} also yields a pair of models, this
time with $SO(2n-1)$ gauge group, while case~(6) in
Figure~\ref{fig:e-e-f} yields a pair of models with $F_4$ gauge 
group. 

Finally, we can also consider case (2),
\begin{equation}
0  \longrightarrow \Gamma(A_{2n}) \longrightarrow \Gamma(D_{2n+3} )
\longrightarrow  \BZ_4 \longrightarrow 0.
\end{equation}
In this case, we see that, on both sides, one of the edges of the
diagram gets identified with itself with the opposite
orientation. This means that one destroys the corresponding Cartan
generator and, moreover, one breaks half of the supersymmetry in the
process. These theories do not have $(1,1)$ supersymmetry in six
dimensions and we do not obtain new matrix models from them. 

The correspondence with the classification of surface quotient
singularities seems to explain why we get a pair of matrix models for
each gauge group. However, several questions remain unanswered. 

The twisted $\Gamma$ and $\Gamma^\prime$ models equivalent both 
appear
to have an interpretation as M-Theory on $(\BC^2\times S^1)/\BZ_n$,
but it would be interesting to see if the six-dimensional gauge
theories have the same or different value of $\theta$. Witten's
analysis in the (degenerate) case~(1) would tend to suggest that the
value of $\theta$ is what distinguishes the pair of models.

Though we presented a reasonable argument for why we find different
rules for extracting the new roots and metric in the twisted $\Gamma$
and $\Gamma^\prime$ models, it would be nice to have a better
understanding. Heuristically, the wrapped membrane states in these
models are associated with the $\BZ_n$-equivariant homology of the 
ALE spaces. An application of the wrapped membrane model discussed in 
section~\ref{sec:wrapped} to these orbifolded theories, as well as a 
construction of the Cartan generators for the $D$ and $E$ series, 
could be used to make this relationship more precise.  

\section{Some Dynamical Considerations}
\label{sec:dynamics}

The ALE space matrix theories have the equivalent of $\CN=2$
supersymmetry in four dimensions. In general, this means that there
are potential one-loop corrections to the metric on their Coulomb
branches, but the standard non-renormalization theorem should protect
against any higher-loop corrections. It is therefore possible that an
$F^2/r^{4-d}$ potential is generated at one-loop between two D0-
branes
which are ground states of the ALE model. This would represent an
interaction between the massless particles that is proportional to 
the
square of their relative 
velocity, which is forbidden if the matrix description is to
properly reproduce the results of supergravity at low
energies. 

The beta function for these theories vanishes, however, as can be 
seen
from the following argument for the $A_{n-1}$ series. The
hyperk\"ahler quotient leaves over $n$ vector multiplets in the
adjoint of the diagonal subgroup 
$U(N)_{\text{diag.}} \subset \times_{i=1}^n U(k_iN)$.
Furthermore, under this subgroup,
$(1,\ldots,1,Nk_i,\overline{Nk}_{i+1},1,\ldots,1)\sim
1+\text{ad}(U(N)_{\text{diag.}})$, so there are also $n$ charged
hypermultiplets in the adjoint. Since the contribution to the beta
function of a vector multiplet will cancel that of a adjoint
hypermultiplet of the same mass, the beta function will be zero if 
the
vector and hypermultiplet masses are paired accordingly.
In the $A_1$ blow-up, it is easy to  check  that, when both states 
are
either in the same or different  vacua, the mass terms do indeed 
match
and the beta function vanishes accordingly.  

The next leading contribution is the $F^4$ interaction, which is
generated at one-loop. In the case of additional compactification on 
a
$d$-torus, the hyperk\"ahler construction goes through
unmodified. There is an integration over the modes of the torus, as 
well as over the compact zero-mode of the gauge
field, yielding a potential which is proportional to $v^4/r^{7-d}$,
as expected from the exchange of gravitons in the infinite
momentum frame~\cite{Berenstein:MatrixVarious}. 
It is  unclear how the degrees of freedom associated to wrapped
membranes might modify this result.

Now in section~\ref{sec:orbifold}, we also gave a definition for 
certain
orbifolds,
M(atrix)[$\CM_{\vec{\zeta}=0}\times S^1/\BZ_n$]. We also find mass 
matching in
this case. Here, the vector and the hypermultiplet will both have
zero mass, but their momenta are quantized in integer and fractional
units respectively, so there might be an overall non-zero
result. However, integrating over the zero-mode of the gauge field
changes the mass terms in the integral so that both contributions
exactly agree. Once again the beta function vanishes and there are no 
$v^2$ 
interactions. We similarly obtain an interaction which is 
proportional
to $v^4/r^6$.
  
We note that, as in the standard matrix SYM, renormalizability will
limit the number of dimensions one can toroidally compactify within
the SYM paradigm. Here, since there are abelian fields coupled to
charged particles, there is sick UV behavior in four or more
dimensions. To provide a sensible definition of the four-dimensional
matrix ALE theory, new degrees of freedom must be added. The simplest
field-theoretic solution is to restore some broken non-abelian
gauge symmetry at some cutoff energy, so that the result is 
consistent.
In any case, the SYM description is only valid up to 
$M[\CM_{\vec{\zeta}=0}\times  T^2]$, which is still short of four 
flat
transverse dimensions.

\section{Conclusions}

The description of massless vectors in the ALE matrix models provides
more evidence that they capture important ingredients of M-Theory on
an ALE space. We have provided evidence that these vectors exist as
normalizable ground states of the Hamiltonian and that they actually
can carry an arbitrary amount of longitudinal momentum by 
establishing their description within matrix string theory. 

We also gave a quantitative prescription for studying wrapped 
membrane states in the ALE matrix theories. Certain properties, such 
as the membrane mass and the membrane-antimembrane Coulomb 
interaction, emerge straightforwardly in our description in the case 
of the $A_1$ singularity. The application of these methods to the 
rest of the $A$-$D$-$E$ cases and more complicated configurations 
than we have considered would yield quite a bit of useful information 
about the ALE matrix models. The masses of these states can be 
computed in a straightforward fashion and they don't receive any 
corrections, as expected from the BPS nature of these states. Of even 
more interest is the structure of the bound states in these 
quantum mechanical systems and the dynamical information that may 
be extracted from them, particularly in the large $N$ limit.  

We have also given, within this framework, an explicit construction
that suggests how orbifolds can be constructed via twists in the
$1+1$-dimensional matrix models. We saw that there were pairs of
twisted matrix models that led to the same gauge groups, yet the 
rules which led to their construction were very different. We 
provided a connection between these pairs of models via Reid's exact
sequences. These twisted matrix models may be the matrix theory
realization of the M-Theory orbifolds on $(\BC^2\times S^1)/\Gamma$,
recently described by Witten~\cite{Witten:NewGauge}.  

\section*{Acknowledgements}

We would like to acknowledge fruitful discussions with Ofer Aharony,
Philip Candelas, Willy Fischler, and Moshe Rozali. We would also like
to thank Edward Witten for a useful correspondence.

We thank Michael Douglas for raising some critical issues regarding 
the discussion of wrapped membrane states and the finite $N$ matrix 
models that appeared in an earlier version of this paper. While we 
were in the process of writing up our revisions to 
section~\ref{sec:wrapped} on wrapped membranes, 
\cite{Diaconescu:Fractional-Branes}~appeared, which also gives a 
quantitative prescription for the study of wrapped membranes in the 
ALE matrix models. 

\renewcommand{\baselinestretch}{1.0} \normalsize

%\nocite{*} %show EVERYTHING in bibliography database (otherwise, get 
%only CITED refs)

\bibliography{strings,m-theory,susy}
\bibliographystyle{utphys}

\end{document}